# Mutualistic Relationships in Service-Oriented Communities and Fractal Social Organizations

Vincenzo De Florio[1], Hong Sun[2], Mohamed Bakhouya[3]
[1]MOSAIC group, Universiteit Antwerpen & iMinds
2020 Antwerpen-Berchem, Belgium
vincenzo.deflorio@uantwerpen.be
[2]AGFA Healthcare, 9000 Gent, Belgium
hong.sun@agfa.com
[3]International University of Rabat
11 100 Sala el Jadida, Morocco
mohamed.bakhouya@uir.ac.ma

***Abstract*— In this paper we consider two social organizations – service-oriented communities and fractal organizations – and discuss how their main characteristics provide an answer to several shortcomings of traditional organizations. In particular, we highlight their ability to tap into the vast basins of "social energy" of our societies. This is done through the establishing of mutualistic relationships among the organizational components. The paper also introduces a mathematical model of said mutualistic processes as well as its translation in terms of semantic service description and matching. Preliminary investigations of the resilience of fractal social organizations are reported. Simulations show that fractal organizations outperform non-fractal organizations and are able to quickly recover from disruptions and changes characterizing dynamic environments.**

***Keywords: Service-oriented communities, fractal organization, agent-based simulations, mutualistic relationships, semantic service description and matching.***

## I. Introduction

With the increase of the human population, resources are becoming scarcer, and a smarter way to make use of them becomes a vital necessity if we want to guarantee welfare and get rid of or at least postpone unmanageability. This is particularly true in world areas where poverty is widespread and primary assets are minimal — including water, food, and medicine. A major problem we observe is that human organizations appear to have been conceived with a different and less challenging context in mind. Though apparently effective when the context was different and a large amount of resources was available to treat a smaller demand, traditional organizations prove now to be too expensive and unable to match the turbulence of the new environments they are set to operate in. Merely expanding the current organizations without properly evolving them is simply not working anymore; and ever more often we are confronted with designs that look like an aqueduct that does indeed distribute its "water", but it does so while losing an unacceptable amount of resources along its way due to leakage in its pipelines. This leakage is often leakage of *social energy* [1] — the intrinsic potential of people and organizations to collaborate and contribute to the common welfare.

We observe that this is due to two major facts. First, traditional organizations introduce predefined roles that divide the agents according to their hypothesized capabilities. A typical example can be found in the domain of healthcare: there, most organizations impose a distinction between professional care-givers (viz., agents able to provide complex, specialized care services); informal care-givers (agents that may occasionally provide non-specialized services); and care-takers (agents that are only on the receiving side of the service chain and are assumed not to be able to provide any contribution to the organization and the other agents). As the majority of the agents lie in the third category, traditional organizations systematically prevent a large amount of social energy to be used for the benefit of our communities.

Secondly, traditional organizations are often structured as "individual-context" systems [2], namely systems incapable of complex inter-organizational collaboration. Evidence of the limitations of the status quo is gathered by observing, e.g., the many deficiencies experienced during the well-known Katrina crisis in New Orleans. As pointed out, e.g., in [3], rescue organizations after Katrina worked in isolation; did not share their knowledge and resources in function of the context; and were incapable of integrating the action of spontaneous responders (so-called "shadow responders"). This brought to a number of collaboration and coordination failures that slowed down the intervention and resulted in significant losses.

The structure of this article is as follows: in Sect. II we recall the main characteristics of two organizational components that – we conjecture – may provide an answer to the above limitations: service-oriented communities (SoC) and fractal social organizations (FSO). In the same section we also show how both SoC and FSO are open social organizations that tap into the vast potential of social energy of our societies. In so doing, SoC and FSO provide a foundation for the definition of "smarter organizations" characterized by greater resilience, scalability, and performance. In order to provide elements of evidence to our conjecture, in this paper we propose a mathematical model of mutualistic relationships. After this in Sect. III we make use of a simulation model to show the enhanced robustness of fractal organizations when dealing with knowledge diffusion. Section IV concludes with a brief view to future investigations.

## II. SERVICE-ORIENTED COMMUNITIES AND THEIR FRACTAL ORGANIZATION

### *A. Service Oriented Community*

Service-oriented Communities (SoC) are social organizations that are designed to tap into the vast basins of social energy of our societies. This is done through two major concepts:

1. The first such concept is that of the member. Instead of casting predefined roles on its components, classifying them into providers and receivers of services, SoC treat all components as peers whose role is defined dynamically considering, e.g., the situation at hand; the competence; the availability; and the service policy of each component. By avoiding artificial classifications into active and passive agents the full potential of our societies can be put to use to serve itself, By doing so a smarter way to optimally recombine the available assets is created, which – we conjecture – may provide an effective tool to respond timely and effectively to needs and changes. Members – for instance the people experiencing a critical situation – are diverse and mobile: diversity enhances a community's adaptability [16] and resilience [17,18] while mobility makes it possible to orchestrate dynamically over a territory.

2. The second key conceptual ingredient of SoC is their ability to establish mutualistic relationships [17] among the members. This is discussed in more detail in the following paragraph.

### *B. Mutualistic Relationships in Service-oriented Communities*

Mutualistic relationships are the basis of intra-species and inter-species collaboration in nature. A typical example is given by symbiosis. A second one, established between the natural kingdoms Animalia and Plantae, is represented in Fig. 1: as an organic function, animals produce carbon dioxide and plants produce oxygen; at the same time, animals require oxygen and plants require carbon dioxide. This natural exchange of services creates a mutually beneficial coupling between animals and plants. Mutualistic relationships may also involve several species and create sort of a mutualistic chain (or mutualistic transitive closure.)

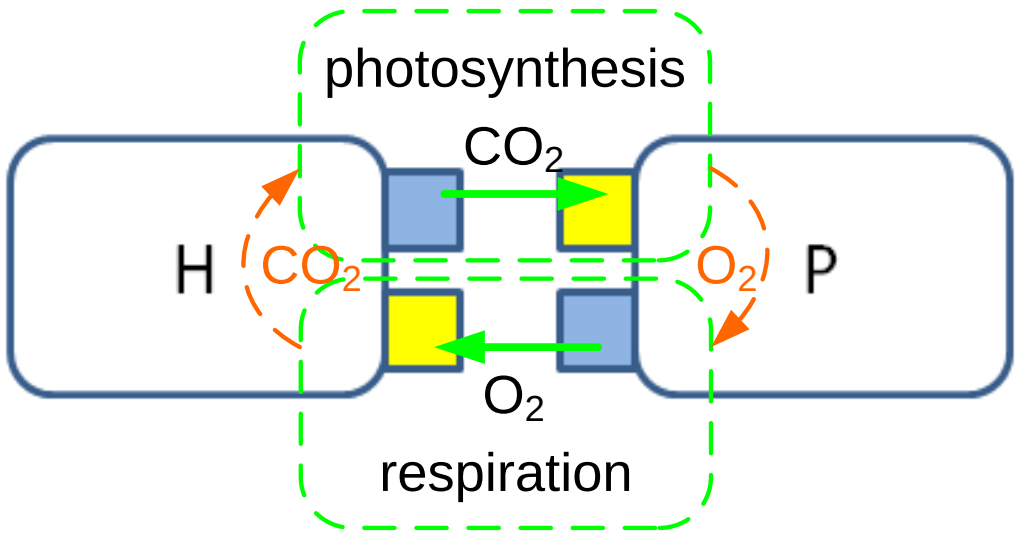

Figure 1. Mutualistic relation between human beings (H) and plants (P).

The second principle of the SoC is *semantic-supported* mutualistic relationships. In SoC, a mutualistic relationship is based on 1) exchange of services triggered by 2) a different "interpretation" of those services. In what follows we introduce the SoC model of mutualistic relationship.

#### *1) Mutualistic Relationship Model*

**Definition II.1** [Action function]. Let D and R be two systems / domains / environments in which actions / events can take place as specified by "action sets" $A_D$ and $A_R$. Then the following bijective function:

$$act: A_D \rightarrow A_R \qquad (1)$$

maps actions in $A_D$ with corresponding actions in $A_R$.

**Definition II.2** [Interpretation/evaluation function]. Let S be a system / domain / environment in which actions / events can take place as specified by action set $A_S$. Then the following function:

$$eval_S: A_S \rightarrow I_S \qquad (2)$$

maps actions in $A_S$ with a semantic interpretation/evaluation of the significance of those actions for S. We assume that said interpretation may be associated at least with one of the following three classes: positive, neutral, and negative, meaning respectively that the mapped action is evaluated as being beneficial, insignificant, or disadvantageous. Integers 1, 0, and -1 will be used to represent the above three classes respectively.

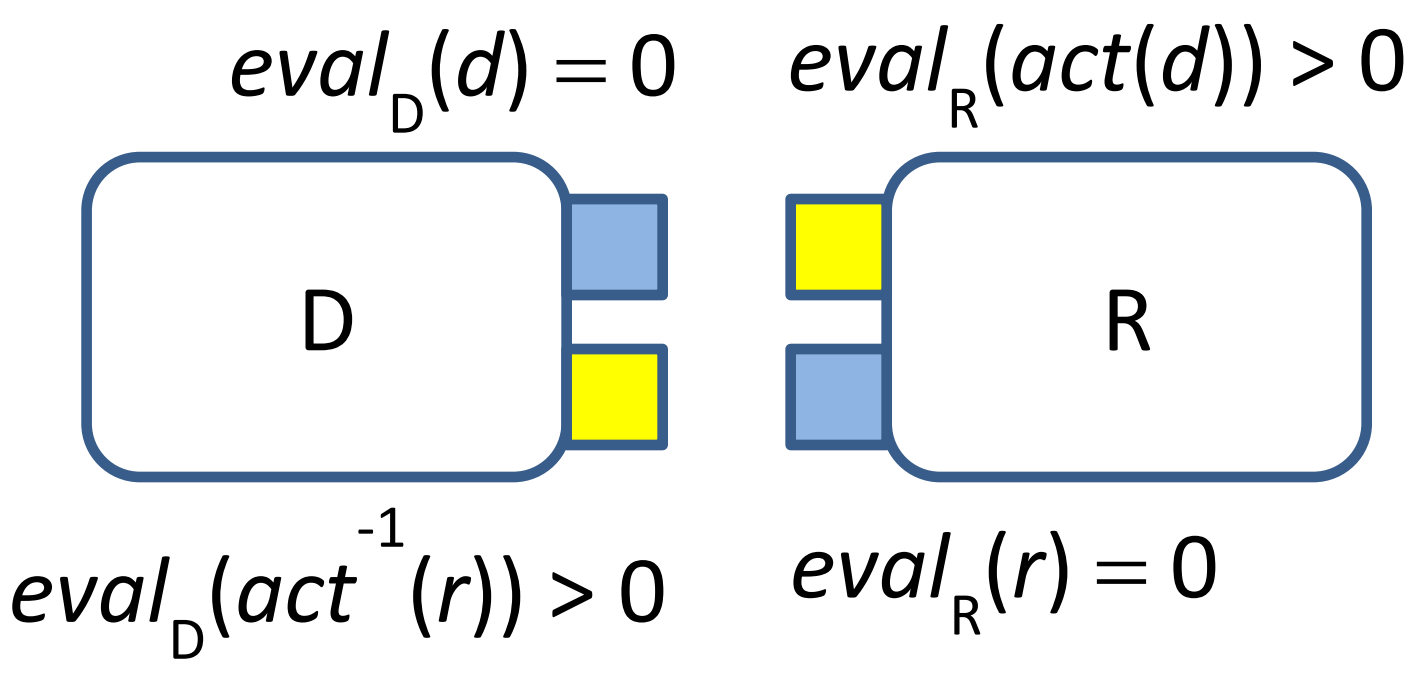

Figure 2. Representation of mutualistic relationship.

We can now define a mutualistic precondition:

**Definition II.3** [Mutualistic precondition]. Let D and R, $A_D$ and $A_R$, and $I_D$ and $I_R$ be defined as above. Then the following conditions are called the mutualistic precondition (MP) between D and R.

$$\exists\, a \in A_D\!: \; eval_D(a) \geq 0 \;\wedge\; eval_R(act(a)) > 0 \qquad (3.1)$$

$$\exists\, b \in A_R\!: \; eval_R(b) \geq 0 \;\wedge\; eval_D(act^{-1}(b)) > 0 \qquad (3.2)$$

The top formula states that an action $a$ in $A_D$ is interpreted as positive or neutral, but its occurrence/consequence in R, $act(a)$, produces positive returns for R. This corresponds to Condition (3.1). The bottom formulae express the dual Condition (3.2): an R-neutral action $b$ in R translates in a beneficial action $act^{-1}(b)$ in D.

**Definition II.4** [Mutualistic relationship]. A mutualistic relationship between two systems / domains / environments D and R is defined as the social behavior occurring when D and R enact individual behaviors that correspond to the mutualistic preconditions (3.1) and (3.2). If D and R are in mutualistic relation we shall write D $\mathscr{R}$ R.

What possibly happens in nature is that the positive returns triggered by a certain D's behavior stimulate in R the production of a dual behavior. The positive interpretation of the latter in D further stimulates the production of the former actions, which consolidates D $\mathscr{R}$ R – namely the mutualistic relationship between D and R.

**Definition II.5** [Extended mutualistic relationship]. Given the same conditions as in Definition II.4, we define as extended mutualistic relationship the social behaviors that correspond to the following conditions:

$$\exists\, a \in A_D:\ eval_R(act(a)) > 0 \qquad (3.3)$$

$$\exists\, b \in A_R:\ eval_D(act^{-1}(b)) > 0 \qquad (3.4)$$

By removing the clauses $eval_R(b) \geq 0$ and $eval_D(a) \geq 0$ we include among the actions triggering a mutualistic relationship also those ones that have a "cost" for the actor – for instance, they require energy consumption. This allows also commercial services to be considered.

*2) Mutualistic Relationships in SoC – towards mutual assistance*

Service-oriented Communities make use of semantic description and matching in order to make it possible for members to establish extended mutualistic relationships. Figure 3 shows that in SoC, a mutualistic relation involves two activities: in activity A1, player 1 provides what player 2 needed (X), while in activity A2, player 2 provides what player 1 needed (Y).

In a SoC, a service description is not mandatory to state what it provides or what it requests. When 'provide' is not stated, this service description is looking for service; when 'request' is not stated, this service description is providing service; when both 'provide' and 'request' are present, this service is looking for a mutualistic relation. It is not allowed to have both 'request' and 'provide' missing.

It is also possible that X and Y are the same service/event. In such a circumstance, the activity is referred to as a group activity, corresponding to what we called as participant model in [11]. In a group activity both parties benefit from participating to the same group activity.

When players are bound together to carry out group activities, a group activity itself can also be considered as a player of the community. For example, if both players in Fig. 3 want to provide and request a *Walking* activity, they would be bound together to take a *Walking* activity. Then the *Walking* activity itself, shown as the green dashed box in Fig. 4, could be considered as a member of the community. The activity would provide a *Walking* activity and request a location. A member 3 who is requesting a *Walking* activity, and a member 4 (for instance, a park), which can provide a location, can be bound with the aforementioned activity; therefore a *Walking* activity which involves members 1, 2, and 3 can be carried out in the location provided by member 4 (i.e., a park). Allowing group activities to act as community players makes it possible to organize activities in SoC as Fractal Social Organizations. In the following section, the concept of Fractal Social Organization will be summarized.

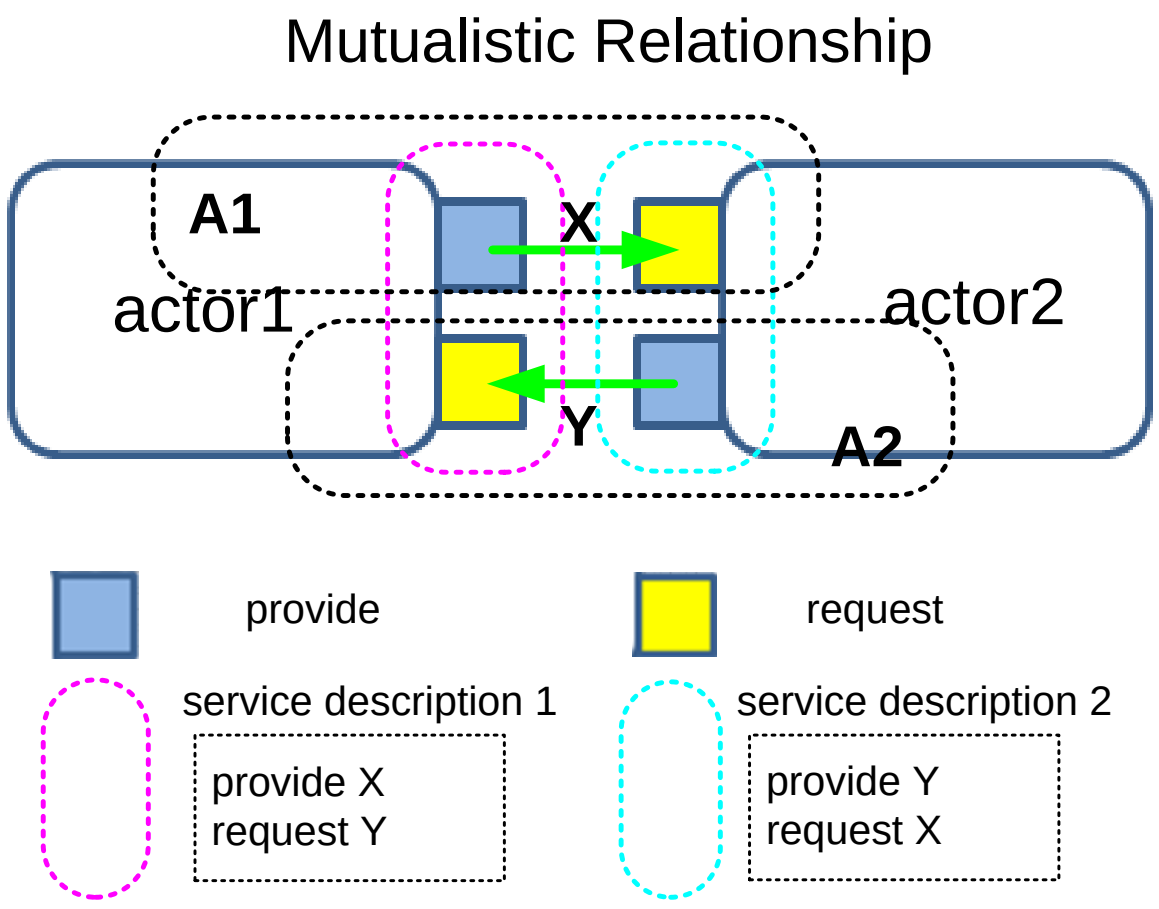

Figure 3. Mutualistic relationship in SoC.

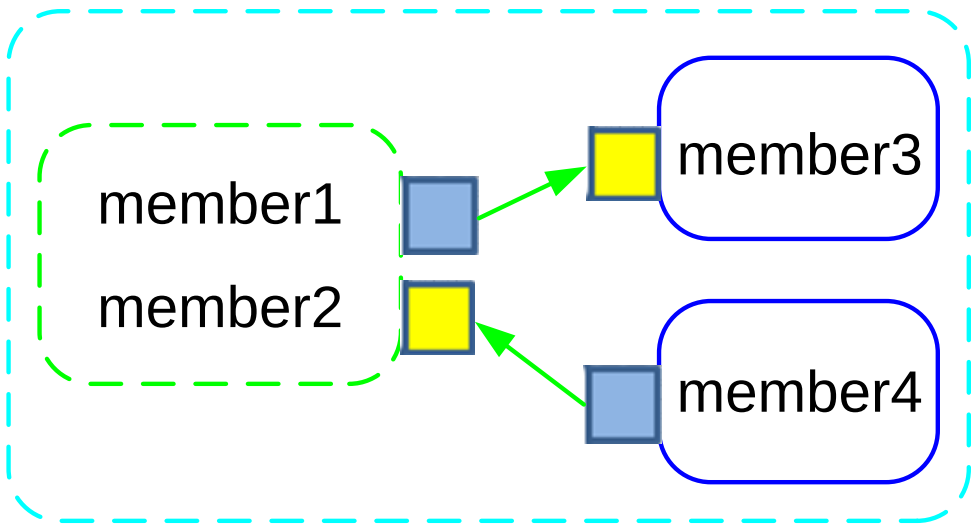

Figure 4. Group activity as individual member.

### C. *Fractal Social Organization*

Fractal Social Organizations are an architecture for socio-technical systems aiming at maximizing Community Resilience, namely "the sustained ability of a community to utilize available resources to respond to, withstand, and recover from adverse situations" [3]. Fractal Social Organizations are Service-oriented Communities whose members can be other Service-oriented Communities. This simple addition translates in a structure resembling that of a "matryoshka doll" and exemplified in Figure 5.

A major difference with respect to "conventional" SoC is in the way actors are enrolled to a servicing protocol. While in an SoC this enrollment involves the members within the SoC, in an FSO enrollment takes place through inter- and intra-SoC collaboration. This is done by linking together members of different SoC by means of the concept of exceptions. Exceptions work as follows: when a triggering condition requires a response activity, the SoC where the condition "fired" tries and locate members to play the roles required by a response activity. If the SoC is able to find all the necessary roles, the response activity is launched. When this is not the case, the SoC triggers a so-called "exception", namely it declares that one or more roles are missing and forwards the condition and the current state to the next level of SoC. By traversing the SoC hierarchy, roles are found and members are assigned to the response activity. So-called “social overlay networks” are thus created, namely temporary SoC whose members have joined from different SoC in order to deal with the condition at hand. The objective and lifespan of the social overlay network are determined by the span of the triggering condition.

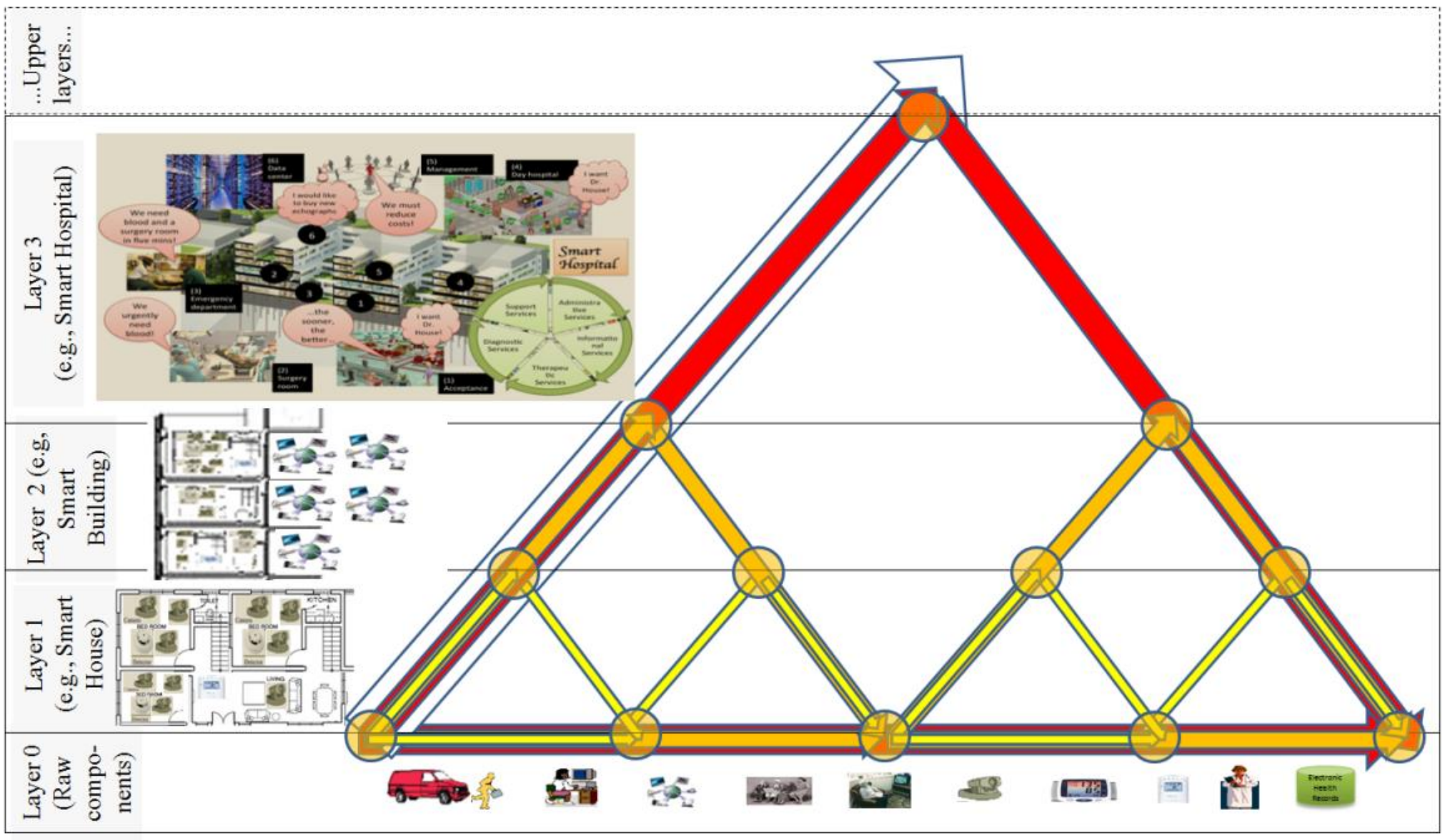

Figure 5. Exemplification of a Fractal Social Organization.

A major advantage of the FSO is that it allows complex inter-organizational collaboration to be modeled in a straightforward way. This includes context-driven resource sharing among organizations; ability to integrate resources both institutional and spontaneous in nature [18]; avoidance of multiple uncoordinated responses that jeopardize the effectiveness of the intervention; avoidance of resource wasting or duplication; avoidance of interventions masking each other out; empowering the communities in “taking charge of their own health” [9]. The result is a “smarter organization” that is able to tap into the vast basins of social energy of our human communities so as to provide a more intelligent way to make optimal use of the available resources; respond to situations and crises; avoid organizational conflicts and resource wasting; and adapt to turbulent or resource-scarce environments.

A system compliant to this model is the software architecture of project Little Sister [5]. Little Sister (LS) aims to **“**research, implement and demonstrate low-cost autonomous technology to provide protection and assistance to elderly citizens” [6]. The LS Service-oriented Communities represent smart flats; smart homes; smart buildings; and smart organizations (e.g., smart hospitals). The LS members are expressly designed low-resolution sensors and RFID readers that are individually wrapped and exposed as manageable web services. The name of the project comes from the fact that the low resolution of the sensors inherently guarantees privacy; thereby avoiding “Big Brother is watching you” syndrome.

These services are then structured within a hierarchical federation reflecting the structure of the community in which they are deployed. More information about the Little Sister software architecture is available at [5]. Further information about Fractal Social Organizations may be found in [10].

## D. *Semantic service description and matching*

SoC instrument a publish-subscribe mechanism between the members of a community. Service descriptions in our proposed SoC are presented following the service descriptions in Fig. 6. In order to bind different service providers and service requesters together, it is important that different service descriptions are using a ‘same language’ which can be understood by each other. Semantic web technology is thus used for service description and matching [15]. As an example, Fig. 6 shows a sample of service description, which indicates that the requested service and provided service is the same (*service:Walking*), which means the service publisher wants to enact *Walking* with others (as a group activity).

```
@prefix service: <http://www.pats.ua.ac.be/AALService#> .
@prefix xsd:    <http://www.w3.org/2001/XMLSchema#> .

[ a                         service:Activity ;
  service:creationTime      "2013-05-12T13:00:00"^^xsd:dateTime ;
  service:endTime           "2013-05-12T21:00:00"^^xsd:dateTime ;
  service:hasCreator        <http://www.pats.ua.ac.be/aal/user/15441#this> ;
  service:hasServiceLocation [ a       <http://schema.org/Beach> ;
                               <http://dbpedia.org/ontology/location>
                                       <http://dbpedia.org/resource/Borgerhout>
                             ] ;
  service:provide           service:Walking ;
  service:request           service:Walking ;
  service:startTime         "2013-05-12T17:00:00"^^xsd:dateTime
] .
```

Figure 6. Sample service description

Once the service descriptions are represented semantically with common ontologies, it is possible to match those literally different but semantically similar services. As an example, the service type *Walking* could be considered as a match to a service requesting for *Fitness* – provided that the user agrees about such inference and at the same time *Walking* is explicitly stated as a sub class of fitness in the ontology. Meanwhile, if a service type *Fitness* is stated in the service description, it can also be deduced that it is possible to provide *Jogging*, *Cycling*, and *Walking* services (provided those services are defined as sub class of *Fitness*). Through such inferencing, the chances to have a service match are much increased. Examples of such inference can be found in [7].

Service publication and service discovery by a Fuseki SPARQL endpoint are experimented in [7]. The published service descriptions are stored in the TDB database (a component of Jena for RDF storage and query) provided by Fuseki. In the reasoning-and-coordination center, requests are served by associating service protocols; identifying required roles; and appointing roles to members while optimizing individual and social concerns. The SoC semantic reasoning allows mutualistic relationships between requests and roles to be identified, which allows, e.g., two requests for service to fulfill each other [4]. Complex reasoning can be carried out by semantic reasoning engines, such as EYE [12]. The SoC model thus enables a self-serve paradigm that promotes active behaviors; stimulates self-management; and helps automating the location and binding of resources. A compliant Web Services-based middleware has been developed and tested in the framework of the already mentioned project LittleSister [5, 6]. Information about the semantic description and matching strategy of the SoC is available, e.g., in [7]. In addition, when an organization is structured as a SoC with activities coordinated by a SPARQL endpoint, it is also possible to connect multiple organizations as FSO by linking those SPARQL endpoints together. A semantic framework which is able to organize a community as FSO is presented in detail in [5].

## III. SIMULATION RESULTS

In this section we have evaluated some of the benefits of a fractal organization with respect to a conventional, hierarchical organization. This has been done using the ORA-based tools developed by CASOS group at Carnegie Mellon for dynamic meta-network assessment and analysis. The ORA tools allow visualizing, assessing, and reasoning about networks (e.g., social and financial networks) [13, 14]. For this initial investigation, we have mainly focused on assessing the resilience capacity of these organizations under dynamically changing conditions, i.e., their ability to recover after isolating one or a set of participating members. We have used two organizations with 15 members each having 15 different knowledge units and each performing a task. The interaction between members or agents define the structure of the network (i.e., who knows who). The figure also shows knowledge units and tasks assigned to each of the participants.

In order to reveal the organizations vulnerabilities the Dynanet tool is used. Dynanet is a dynamic network analysis and Near-Term analysis tool based on agent-based paradigm able to simulate member removals during the simulation and get various performance metrics, such as knowledge/information diffusion. The first investigation was to evaluate the performance of these two organizations under various isolation strategies. More precisely, we have conducted simulations and used information diffusion measure to figure out the performance changes for the organizations. In other words, we have injected changes at different points in time to figure out the resilience capability of these organizations.

Three scenarios are considered. The first scenario is shown in Fig. 7, in which no destabilization is performed. As depicted in that figure, the diffusion rate goes up as simulation progress. The fractal organization shows the highest information diffusion rate. Furthermore, the diffusion rate converges quickly and smoothly. Figure 8 illustrates the second scenario in which one member is isolated at time 10. As depicted in that figure, at time 10, the isolation of one member translates into a somewhat damaged diffusion rate compared to the first scenario, for both organizations; both of them are able to recover in the following time periods, but the fractal organization shows highest information diffusion rate. Figure 9 illustrates a third scenario in which multiple members are isolated (5 in this case). The first isolation takes place at time 10, and the next isolations happens after a gap of 10, 20, 30, and 50 time periods, respectively. Clearly in both cases the isolation of the five members prevents full information diffusion. The rate is lower for both organizations when compared to the first and second scenarios, but still the fractal organization recovers better.

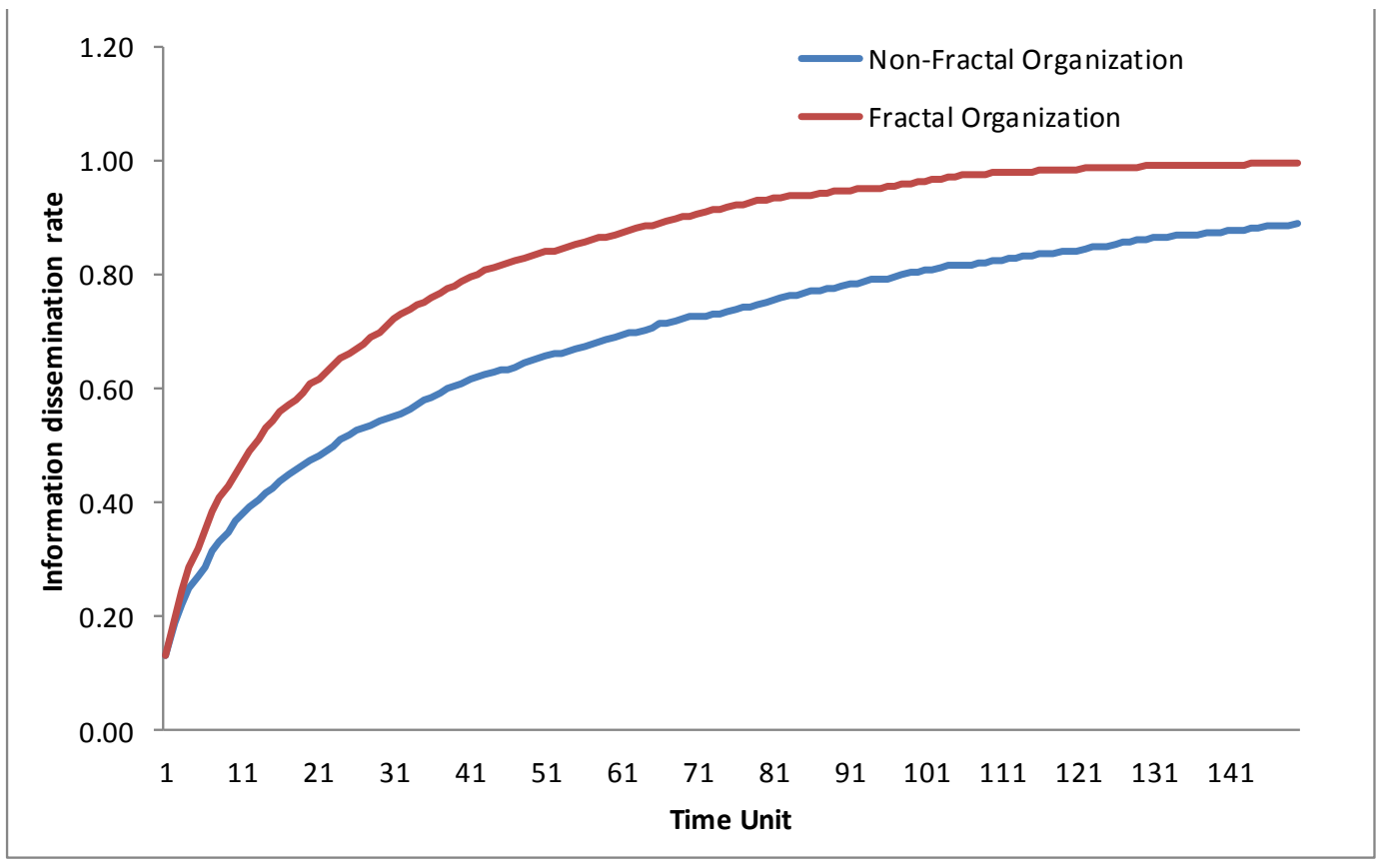

Figure 7. Baseline scenario. No isolation.

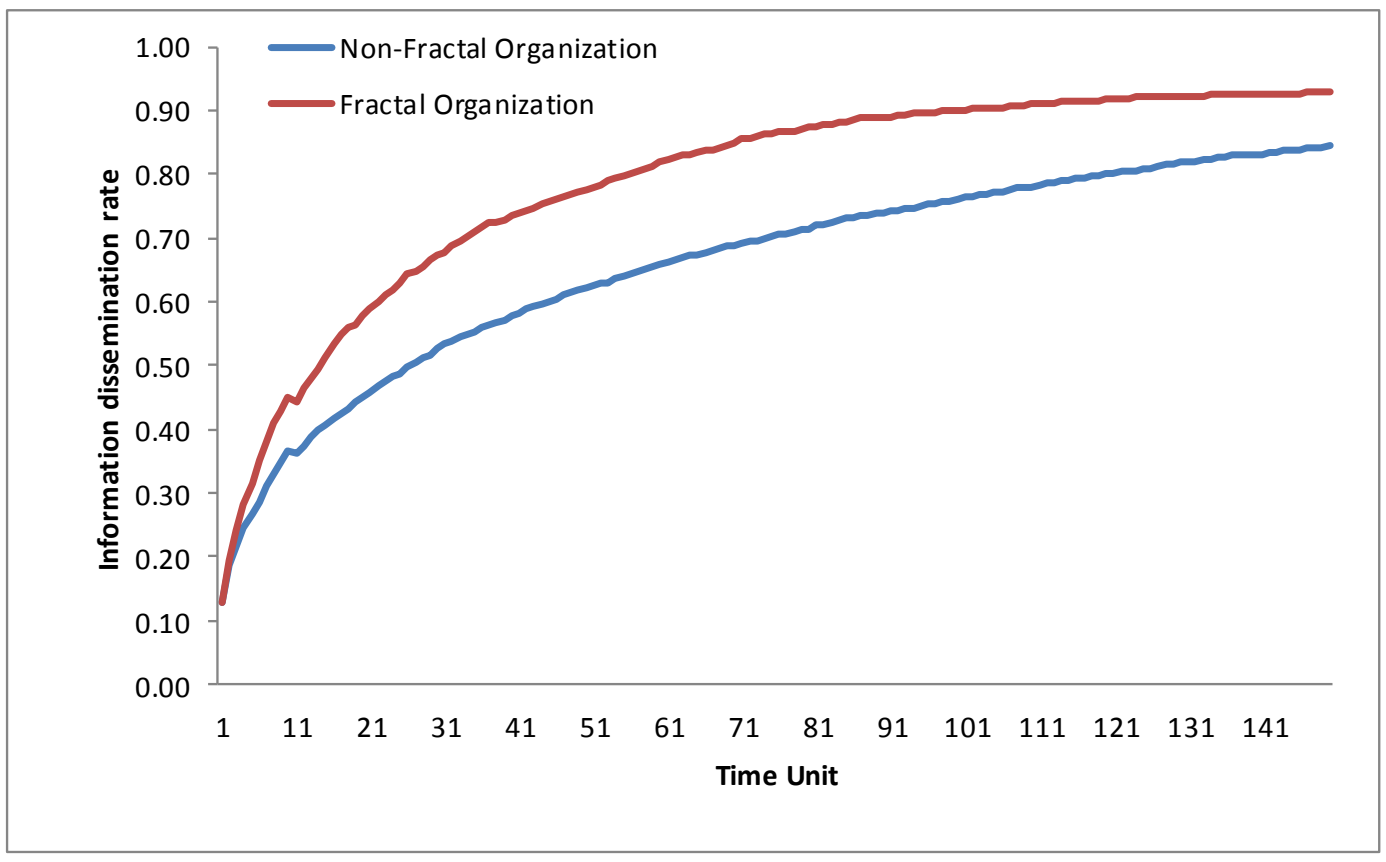

Figure 8. One member isolation scenario.

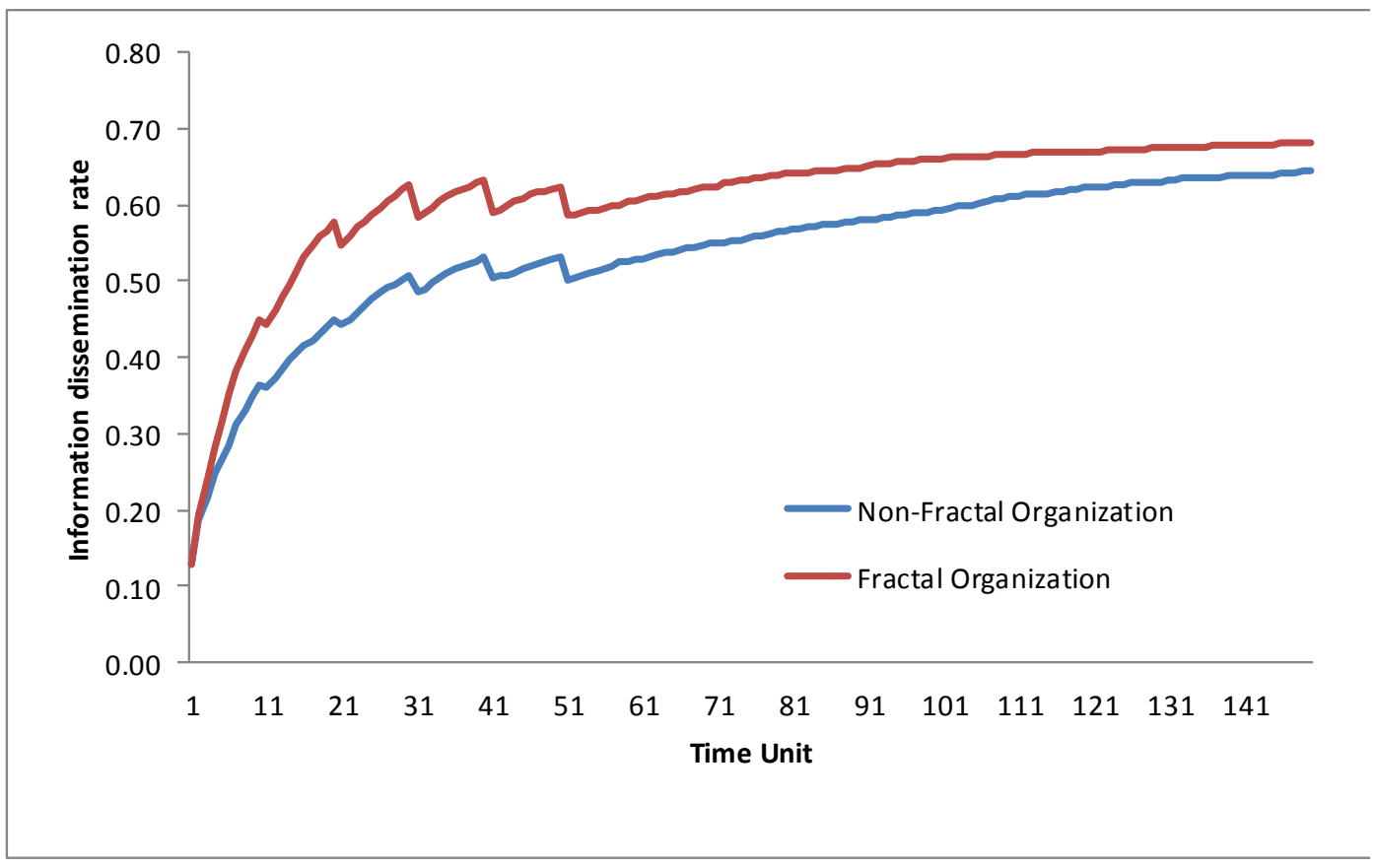

Figure 9. Five member isolation scenarios.

## IV. Conclusions and perspectives

In this paper we have suggested how the concepts of service-oriented communities and fractal social organizations may provide our societies with an alternative organization for important services such as care and crisis management. In particular, we suggested the role of fractal social organizations as socio-technical systems for the engineering of community resilience services. We have shown the potential of mutualistic approaches through a mathematical model; provided examples of how this model translates in terms of semantic service description and matching; and shown how a fractal organization appears to outperform traditional organizations in terms of resilience. In particular fractal organizations are able to quickly recover because isolating members does not leave the rest of members disconnected. In our ongoing work we are using the concept of service-oriented community to study how fractal organizations could heal and recover against isolation strategies. Our main aim is to study that capability and how faster fractal organizations could be restructured against dynamic and unexpected changes.

## References

[1] De Florio, Vincenzo and Blondia, Chris (2010). Service-Oriented Communities: Visions and Contributions towards Social Organizations. On the Move to Meaningful Internet Systems: OTM 2010 Workshops. LNCS 6428. Springer.

[2] Eugster, Patrick Th., Garbinato, Benoit, and Holzer, Adrian (2009). Middleware support for context aware applications. Chapter 14 in "Middleware for Network Eccentric and Mobile Applications", pp. 305-322. Springer.

[3] Colten, C. E., Kates, R.W., and Laska, S.B. (2008). Community Resilience: Lessons from New Orleans and Hurricane Katrina. CARRI Research Report 3. (Sept. 2008). Online: http://www.rand.org/topics/community-resilience.html.

[4] Sun, Hong, De Florio, Vincenzo, Gui, Ning and Blondia, Chris (2010). The Missing Ones: Key Ingredients Towards Effective Ambient Assisted Living Systems. Journal of Ambient Intelligence and Smart Environments **2**(2), pp. 109-120.

[5] De Florio, Vincenzo, Sun, Hong, Buys, Jonas, and Blondia, Chris (2013). On the Impact of Fractal Organization on the Performance of Socio-technical Systems. Proc. of the 2013 Int.l Workshop on Intelligent Techniques for Ubiquitous Systems (ITUS 2013), pp. 594-607. IEEE.

[6] Little Sister: Project description. Online: http://www.iminds.be/en/research/overview-projects/p/detail/littlesister.

[7] Sun, Hong, De Florio, Vincenzo, and Blondia, Chris (2013). Implementing a Role Based Mutual Assistance Community with Semantic Service Description and Matching. Proc. of the Int.l Conference on Management of Emergent Digital EcoSystems (MEDES). ACM.

[8] Buck, John and Endenburg, Gerard (2012). The Creative Forces of Self-organization. Technical Report, Sociocratic Center, Rotterdam, The Netherlands. Online: http://www.governancealive.com/wp-content/uploads/2009/12/CreativeForces_9-2012_web.pdf

[9] Anonymous (2010). Evaluation Report of the Community Health Strategy Implementation in Kenia. Online: http://www.unicef.org/evaldatabase/files/14_2010_HE_002_Community_Strategy_Evaluation_report_October_2010.pdf

[10] De Florio, V., Bakhouya, M., Coronato, A., Di Marzo G. (2013). Models and Concepts for Socio-technical Complex Systems: Towards Fractal Social Organizations. Sys Res and Behav Sci **30**(6). Wiley.

[11] Sun, H., De Florio, V., Gui, N., and Blondia, C. (2007, June). Participant: A New Concept for Optimally Assisting the Elder People. In Proc. of the 20th IEEE Int.l Symposium on Computer-Based Medical Systems (pp. 295-300). IEEE Computer Society.

[12] Euler: Yet another proof Engine. Online: http://eulersharp.sourceforge.net/2003/03swap/eye-note.txt

[13] Carley, Kathleen M., Pfeffer, Jürgen, Reminga, Jeffrey, Storrick, Jon, Columbus, Dave. ORA User's Guide 2013. CMU, Inst. for Software Research, Technical Report, CMU-ISR-13-108.

[14] Carley, Kathleen M., forthcoming, "Dynamic Network Analysis" in the Summary of the NRC workshop on Social Network Modeling and Analysis, Ron Breiger and Kathleen M. Carley (Eds.), National Research Council.

[15] Bakhouya, M., Gaber, J. (2007). "Service Composition Approaches for Ubiquitous and Pervasive Computing Environments: A Survey" In: Agent Systems in Electronic Business, (Eldon Li and Soe-Tsyr Yuan, Eds.), pp. 323-350, Information Science Reference/IGI Publishing.

[16] Stark, David C. (1999). Heterarchy: Distributing Authority and Organizing Diversity. In "The Biology of Business: Decoding the Natural Laws of Enterprise," J.H. Clippinger III (Ed.), pp. 153-179, Jossey-Bass.

[17] De Florio, Vincenzo (2014). Quality Indicators for Collective Systems Resilience. ArXiv e-prints. Online: http://adsabs.harvard.edu/abs/2014arXiv1401.5607D.

[18] De Florio, V., Sun, H., and Blondia, C. (2014). Community Resilience Engineering: Reflections and Preliminary Contributions. SERENE 2014, LNCS 8785, pp. 1–8, Springer.